\documentstyle[preprint,aps]{revtex}

\begin{document}
\draft
\title{Energy and angular momentum of charged rotating black holes}
\author{J.~M.~Aguirregabiria, A.~Chamorro and
K.~S.~Virbhadra\thanks{Present adress: Tata Institute of Fundamental
Research, Homi Bhabha Road, Bombay 400005, India.}}
\address{Dpto. de F\'{\i}sica Te\'{o}rica, Universidad del Pa\'{\i}s
Vasco, Apdo.~644, 48080 Bilbao, Spain }
\maketitle
 \begin{abstract}
 We show  that the pseudotensors
of Einstein, Tolman, Landau and Lifshitz, Papapetrou, and Weinberg
(ETLLPW) give the same distributions of energy, linear momentum and
angular momentum, for any Kerr-Schild metric. This
result generalizes a previous work by G\"urses and G\"ursey that dealt
only with the pseudotensors of Einstein and Landau and Lifshitz.
We compute these distributions for the Kerr-Newman and Bonnor-Vaidya
metrics and find reasonable results.
All calculations are performed
without any approximation in  Kerr-Schild Cartesian coordinates.
For the Reissner-Nordstr\"{o}m  metric these definitions give the same
result as the Penrose quasi-local mass. For the Kerr black hole the
entire energy is confined to its interior whereas for the Kerr-Newman
black hole, as expected, the energy is shared by its interior as well
as exterior. The total energy and angular momentum of the Kerr-Newman
black hole are $M$ and $ M a$, respectively ($M$ is the mass parameter
and $a$ is the rotation parameter). The energy distribution for the
Bonnor-Vaidya metric is the same as the Penrose quasi-local mass
obtained by Tod.
 \end{abstract}
\section{Introduction}
\label{sec:intro}

The general theory of relativity is an excellent theory of space, time
and gravitation and has been  supported by experimental evidences with
flying colors, but some of its features are not without  difficulties.
For instance, the subject of energy-momentum localization has been a
problematic issue since the outset  of this theory. Einstein
investigated whether or not one can obtain a locally conserved
energy-momentum tensor for the gravitational field plus the source
(given by the right hand side of the Einstein equations). However, the
locally conserved energy-momentum complex constructed by him is
neither a tensor nor it is symmetric and therefore its physical
interpretation was questioned by several physicists, notably by Weyl,
Pauli, and Eddington (see reference \cite{ref1}). Tolman\cite{ref2}
obtained a new energy-momentum pseudotensor which is again not
symmetric. However, Landau and Lifshitz (LL)\cite{ref3} succeeded in
constructing a symmetric energy-momentum pseudotensor which can
therefore be used to obtain the angular momentum of a general
relativistic system. Nevertheless, to use the pseudotensors of
Einstein, Tolman, or LL, one is restricted to quasi-Minkowskian
coordinates. M{\o}ller\cite{ref4}, arguing  that to single out a
particular coordinate system is not satisfactory from the general
relativistic point of view, constructed a new energy-momentum
pseudotensor and claimed that with it one was not constrained to use
asymptotically Minkowskian coordinates.  The energy and energy current
density components of the M{\o}ller pseudotensor transform as a
four-vector density with respect to the group of purely spatial
transformations. However, three years later, M{\o}ller  observed a
serious drawback of his prescription\cite{ref5}, i.e., the total
energy-momentum vector of a closed physical system is not a Lorentz
four-vector. Thus, M{\o}ller's attempt to give a
coordinate-independent prescription for energy calculations failed and
therefore we will not discuss M{\o}ller's pseudotensor any more in
this paper. In fact, following the energy-momentum pseudotensor of
Einstein, a plethora of definitions for energy, momentum, and angular
momentum of a general relativistic system have been proposed by many
authors (see \cite{ref6} and references therein). Komar\cite{ref7}
gave a coordinate-independent definition for the energy. Using his
prescription, Cohen and de Felice\cite{ref8} calculated the effective
mass of the Kerr-Newman (KN) metric. The Komar mass for the
Reissner-Nordstr\"{o}m (RN) metric is $E =  M - Q^2/r$ ($M$ and $Q$
are the mass and charge parameters, respectively), which is not in
agreement with the linear theory. Moreover, Tamburino and
Winicour\cite{ref9} pointed out that the Komar definition is not
adequate for radiating systems.
Penrose\cite{ref11} proposed a quasi-local definition of mass,
momentum, and angular momentum in general relativity. Using the
Penrose definition, Tod\cite{ref12} calculated the quasi-local mass
for several spacetimes. For the RN  metric he found $E = M - Q^2/ (2
r)$. He pointed out that as opposed to the Komar energy his result  is
in agreement with the linear theory. However, the Penrose definition
has not  succeeded to deal with the Kerr metric\cite{ref13}.
Bergqvist\cite{ref10} considered seven
different  definitions of quasi-local mass and found
that not any two of them
give the same result for the RN  and Kerr spacetimes.
Despite
these problems there has been considerable interest in this subject in
recent years (see \cite{ref14,refCo} and references therein).

As the energy-momentum complexes of Einstein, Tolman, and LL are not
tensors under general coordinate transformations, many physicists do
not take them seriously as prescriptions for energy-momentum
localization in general relativity. By contrast, the total energy,
momentum, and angular momentum (in LL prescription) are accepted
unanimously when calculations are carried out in quasi-Minkowskian
coordinates. Lindquist, Schwartz, and Misner\cite{ref15}, using the LL
pseudotensor, calculated the energy, momentum, and power output for
the Vaidya metric and got the expected result. One of the present
authors (Virbhadra,  referred to as KSV hereafter)\cite{ref16} showed
that the pseudotensors of Einstein, Tolman, and LL (ETLL) give the
same and reasonable energy distribution in the KN field when
calculations are carried out in Kerr-Schild (KS) Cartesian
coordinates. He also obtained the angular momentum distribution in the
LL prescription. However, his calculations were limited up to the
third order of the rotation parameter. Switching off the charge
parameter he found that there is no energy associated with the
exterior of the Kerr black hole. Though the investigations were
limited up to the third order of the rotation parameter, he
conjectured that one would get the same result for the Kerr metric  if
the calculations were carried out exactly. Cooperstock and
Richardson\cite{ref17} extended the energy calculations up to the
seventh order of the rotation parameter and found that the
pseudotensors of ETLL give the same energy distribution for the KN
metric. Moreover, their result supported the conjecture of KSV that
there is no energy associated with the exterior of the Kerr black
hole. Later on KSV\cite{ref18} showed that the pseudotensors of ETLL
yield the same energy and  energy current density components for the
Vaidya metric.

Recently two of the present authors (Chamorro and KSV)\cite{ref19}
obtained the energy distribution in the Bonnor-Vaidya (BV)
spacetime\cite{ref20} in the prescriptions of Einstein and LL. Both
definitions give the same result as the Penrose
prescription\cite{ref13}. They also obtained the energy current
density components (and power output) for the same metric. Both
(Einstein and LL) prescriptions give the same reasonable result.
Despite these successes this subject required more study. For
instance, there are other pseudotensors known in the  literature and
many more can be constructed (with the property of divergence-free
relation), which could give different results for the KN, BV or other
spacetimes. Moreover, the result known for the KN metric was limited
up to the seventh order of the rotation parameter and it could be
possible that different pseudotensors disagree if calculations were
exactly performed. The aim of this paper is to clarify these
questions. We consider two more well-known (symmetric) energy-momentum
pseudotensors, i.e., the pseudotensors of Papapetrou and
Weinberg\cite{ref21} and show  that all these pseudotensors lead to
the same result for the KN as well as the BV spacetimes when
calculations are carried out in KS Cartesian coordinates. Weinberg,
using his pseudotensor, calculated the total energy, momentum, and
angular momentum of the Kerr metric. He carried out calculations at
infinite radial distance and therefore his results do not bear on
energy-momentum distributions.

Only recently has been brought to our attention that in an interesting
paper G\"urses and G\"ursey\cite{refgg} showed that the pseusotensors
of Einstein and LL coincide for all Kerr-Schild metrics. In this paper
we extend that result by showing that the five pseudotensors of ETLLPW
coincide for any Kerr-Schild metric and, in consequence, give the same
energy and energy current density components for the KN as well as BV
spacetimes. The rest of the  paper is organized as follows: Sec.\
\ref{sec:pseudo} presents the energy-momentum pseudotensors of ETLLPW.
In Sec.\ \ref{sec:ks} we show that the five pseudotensors
coincide if KS Cartesian coordinates can be used. Sec.\
\ref{sec:metric} gives the results for the energy, momentum, and
angular momentum distributions of the KN metric in KS Cartesian
coordinates. The energy, momentum, and  energy current density
components are also given. We present the results corresponding to the
BV metric in Sec.\ \ref{sec:BV}. Sec.\ \ref{sec:discussion} discusses
the results obtained in previous sections.

{\it Conventions.} We use geometrized units in which the speed of
light in vacuum $c$ and the Newtonian gravitational constant $G$ are
taken to be equal to 1, the metric has signature $+ - - -$, and Latin
(Greek)  indices take values $0\ldots3$ ($1\ldots3$).

\section{Energy-momentum pseudotensors}
\label{sec:pseudo}

The energy-momentum pseudotensors of ETLLPW are given below:

(a) The pseudotensor of Einstein is\cite{ref4}
 \begin{equation}
\Theta_i{}^{k} = \frac{1}{16 \pi} {H_i{}^{kl}}_{,l},
\label{eq1}
 \end{equation}
where
 \begin{equation}
{H_i{}^{kl}} =- {H_i{}^{lk}}=
\frac{g_{in}}{\sqrt{-g}}
\left[ -g \left(g^{kn} g^{lm} - g^{ln} g^{km} \right) \right]_{,m}.
\label{eq2}
 \end{equation}
$\Theta_0{}^0$, $\Theta_{\alpha}{}^0$, and $\Theta_0{}^{\alpha}$
are the energy, momentum, and energy current density components.
$\Theta_i{}^k$ satisfies the local conservation laws:
 \begin{equation}
\frac{\partial \Theta_i{}^{k}}{\partial  x^k}=0.
\label{eq3}
 \end{equation}
The energy and momentum  are given by
 \begin{equation}
P_i=\int\!\!\!\int\!\!\!\int{\Theta_i{}^{0}\,dx^1\,dx^2\,dx^3}.
\label{eq4}
 \end{equation}
Using Gauss's theorem one can write
 \begin{equation}
P_i=\frac{1}{16 \pi} \int\!\!\!\int{H_i{}^{0\alpha} n_{\alpha}\,dS},
\label{eq5}
 \end{equation}
where $n_{\alpha}$ is the outward unit normal vector and $dS$ is the
infinitesimal surface element.

(b) The pseudotensor of Tolman is\cite{ref2}
 \begin{equation}
{\cal{T}}_i{}^k = \frac{1}{8 \pi}{U_i{}^{kl}}_{,l},
\label{eq6}
 \end{equation}
where
 \begin{equation}
U_i{}^{kl}=\sqrt{-g}\left[-g^{pk} V_{ip}{}^l +
\frac{1}{2} g^k_{i} g^{pm} V_{pm}{}^l\right],
\label{eq7}
 \end{equation}
with
 \begin{equation}
V_{jk}{}^i = -\Gamma^i_{jk}
               +\frac{1}{2} g^i_{j} \Gamma^m_{mk}
               +\frac{1}{2} g^i_{k} \Gamma^m_{mj}.
\label{eq8}
 \end{equation}

${\cal{T}}_0{}^0$, ${\cal{T}}_{\alpha}{}^0$, and
${\cal{T}}_0{}^{\alpha}$
are the energy, momentum, and energy current density components.
${\cal{T}}_i{}^k$ satisfies the local conservation laws:
 \begin{equation}
\frac{\partial {\cal{T}}_i{}^{k}}{\partial  x^k}=0.
\label{eq9}
 \end{equation}
The energy and momentum  are given by
 \begin{equation}
P_i=\int\!\!\!\int\!\!\!\int{{\cal{T}}_i{}^{0}\,dx^1\,dx^2\,dx^3}.
\label{eq10}
 \end{equation}
For time-independent metrics  one can write
 \begin{equation}
P_i=\frac{1}{8 \pi} \int\!\!\!\int{U_i{}^{0 \alpha} n_{\alpha}\,dS}.
\label{eq11}
 \end{equation}

(c) The symmetric pseudotensor of Landau and Lifshitz is\cite{ref3}
 \begin{equation}
L^{ik}=  \frac{1}{16 \pi} {{\lambda}^{iklm}}_{,lm},
\label{eq12}
 \end{equation}
where
 \begin{equation}
{\lambda}^{iklm}=-g \left(g^{ik} g^{lm}-g^{il} g^{km}\right).
\label{eq13}
 \end{equation}
$L^{00}$ and $ L^{\alpha 0}$  are the energy and energy current
(momentum) density components. $L^{ik}$ satisfies the local
conservation laws:
 \begin{equation}
\frac{\partial L^{ik}}{\partial  x^k}=0.
\label{eq14}
 \end{equation}
The energy and momentum  are given by
 \begin{equation}
P^i=\int\!\!\!\int\!\!\!\int{L^{i0}\,dx^1\,dx^2\,dx^3}
\label{eq15}
 \end{equation}
and the angular momentum is given by
 \begin{equation}
J^{ik}=\int\!\!\!\int\!\!\!\int{\left(x^i L^{0k} - x^k L^{0i}\right)
\,dx^1\,dx^2\,dx^3}.
\label{eq16}
 \end{equation}
Using Gauss's theorem, the energy and momentum are
 \begin{equation}
P^i=\frac{1}{16 \pi} \int\!\!\!\int{ {{\lambda}^{i0\alpha m}}_{,m}
n_{\alpha}\,dS}
\label{eq17}
 \end{equation}
and the physically interesting components
of $J^{ik}$ are
 \begin{equation}
J^{\alpha \beta}=\frac{1}{16 \pi} \int\!\!\!\int{
\left(x^{\alpha}{{\lambda}^{ \beta0 \sigma m}}_{,m}
-x^{\beta}{{\lambda}^{ \alpha0 \sigma m}}_{,m}
+\lambda^{ \alpha 0 \sigma\beta}
\right)n_{\sigma}\,dS}.
\label{eq18}
 \end{equation}

(d) The symmetric pseudotensor of Papapetrou is\cite{ref21}
 \begin{equation}
\Sigma^{ik}=\frac{1}{16 \pi}{N^{iklm}}_{,lm},
\label{eq19}
 \end{equation}
where
 \begin{equation}
N^{iklm}=\sqrt{-g} \left(g^{ik} \eta^{lm} - g^{il} \eta^{km}
+ g^{lm} \eta^{ik} - g^{lk} \eta^{im}\right),
\label{eq20}
 \end{equation}
with
 \begin{equation}
\eta^{ik}= {\rm diag}(1,-1,-1,-1).
\label{eq21}
 \end{equation}

$\Sigma^{00}$ and $\Sigma^{\alpha 0}$  are the energy, and energy
current (momentum) density components. $\Sigma^{ik}$ satisfies the
local conservation laws:
 \begin{equation}
\frac{\partial \Sigma^{ik}}{\partial  x^k}=0.
\label{eq22}
 \end{equation}
The energy and momentum  are given by
 \begin{equation}
P^i=\int\!\!\!\int\!\!\!\int{\Sigma^{i0}\,dx^1\,dx^2\,dx^3}
\label{eq23}
 \end{equation}
and the angular momentum is given by
 \begin{equation}
J^{ik}=\int\!\!\!\int\!\!\!\int{\left(x^i\Sigma^{0k}-x^k\Sigma^{0i}
\right)\,dx^1\,dx^2\,dx^3}.
\label{eq24}
 \end{equation}
For time-independent metrics, one has
 \begin{equation}
P^i=\frac{1}{16 \pi} \int\!\!\!\int{{N^{i0\alpha\beta}}_{,\beta}
n_{\alpha}\,dS}
\label{eq25}
 \end{equation}
and the physically interesting components of the angular momentum are
 \begin{equation}
J^{\alpha \beta}=\frac{1}{16 \pi} \int\int{
\left(x^{\alpha} {N^{0\beta\gamma\sigma}}_{,\gamma}
-x^{\beta} {N^{0\alpha\gamma\sigma}}_{,\gamma}
- N^{0\beta\sigma\alpha}
+ N^{0\alpha\sigma\beta}
\right)
 n_{\sigma}\, dS}.
\label{eq26}
 \end{equation}

(e) The symmetric pseudotensor of Weinberg is\cite{ref21}
 \begin{equation}
W^{ik}= \frac{1}{16 \pi} {D^{lik}}_{,l},
\label{eq27}
 \end{equation}
where
 \begin{equation}
D^{lik}= \frac{\partial h^a_ a}{\partial x_l}\eta^{ik}
         -  \frac{\partial h^a_ a}{\partial x_i}\eta^{lk}
         - \frac{\partial h^{al}}{\partial x^a}\eta^{ik}
         + \frac{\partial h^{ai}}{\partial x^a}\eta^{lk}
         + \frac{\partial h^{lk}}{\partial x_i}
         - \frac{\partial h^{ik}}{\partial x_l}
\label{eq28}
 \end{equation}
and
 \begin{equation}
h_{ik}=g_{ik} - \eta_{ik}.
\label{eq29}
 \end{equation}
$\eta_{ik}$ is the Minkowski metric. Indices on $h_{ik}$ or $\partial/
\partial x_i$ are raised or lowered with help of $\eta$'s.
It is clear that
 \begin{equation}
D^{lik}=- D^{ilk}.
\label{eq30}
 \end{equation}

$W^{00}$ and $W^{\alpha 0}$  are the energy and energy current
(momentum) density components. $W^{ik}$ satisfies the local
conservation laws:
 \begin{equation}
\frac{\partial W^{ik}}{\partial  x^k}=0.
\label{eq31}
 \end{equation}
The energy and momentum  are given by
 \begin{equation}
P^i=\int\!\!\!\int\!\!\!\int{ W^{i0}\, dx^1 \, dx^2 \, dx^3}
\label{eq32}
 \end{equation}
and the angular momentum is given by
 \begin{equation}
J^{ik}=\int\!\!\!\int\!\!\!\int{\left(x^i W^{0k} - x^k W^{0i}\right)
\, dx^1 \, dx^2
\,dx^3}.
\label{eq33}
 \end{equation}
Using Gauss's theorem, one has
 \begin{equation}
P^i=\frac{1}{16 \pi} \int\!\!\!\int{ D^{\alpha 0 i} n_{\alpha}
\, dS}
\label{eq34}
 \end{equation}
and the physically interesting components of the angular momentum are
 \begin{equation}
J^{\alpha \beta}=\frac{1}{16 \pi} \int\!\!\!\int{
\left(x^{\alpha} D^{\sigma 0 \beta}
-x^{\beta} D^{\sigma 0 \alpha}
+\eta^{\sigma\alpha} h^{0\beta}
- \eta^{\sigma\beta} h^{0\alpha}
\right)
 n_{\sigma}\, dS}.
\label{eq35}
 \end{equation}

\section{Kerr-Schild metrics}
\label{sec:ks}

In the following we shall consider the algebraically special metrics
of Kerr-Schild which are given by
 \begin{equation}
g_{ik} = \eta_{ik}-2Vl_il_k
\label{eqks1}
 \end{equation}
in terms of the scalar function $V$ and the null vector $l_i$ which
satisfies the following properties:
 \begin{equation}
g_{ik}l^il^k = \eta_{ik}l^il^k=0,\qquad l^il_{k;i} = l^il_{k,i} = 0.
\label{eqks2}
 \end{equation}
For all these metrics one has $g=-1$, $l^i = g^{ik}
l_k = \eta^{ik} l_k$ and the inverse metric is
 \begin{equation}
g^{ik} = \eta^{ik}+2Vl^il^k.
\label{eqks3}
 \end{equation}
G\"urses and G\"ursey\cite{refgg} pointed out that for these metrics
the
pseudotensors of Einstein and LL coincide and are proportional to the
Einstein tensor:
 \begin{equation}
\Theta_i{}^{k}  = \eta_{ij} L^{jk}= \frac{1}{8\pi} G_i{}^k.
\label{eqks4}
 \end{equation}

In fact,
by using the properties of Kerr-Schild
metrics it is not difficult to prove that for these metrics the five
pseusotensors of ETLLPW will coincide because
one always
have in KS coordinates:
 \begin{equation}
\Theta_i{}^{k} =  {\cal{T}}_i{}^k = \eta_{ij} L^{jk} =
\frac{1}{8\pi}G_i{}^k,\qquad
L^{ik} = \Sigma^{ik} = W^{ik} =
\frac{1}{16\pi}\Lambda^{iklm}{}_{,lm},
\label{eqks5}
 \end{equation}
where
 \begin{equation}
\Lambda^{ikpq}=2V\left(
\eta^{ik}l^pl^q+\eta^{pq}l^il^k-\eta^{ip}l^kl^q-\eta^{kq}l^il^p
\right).
\label{eqks6}
 \end{equation}
In consequence, the energy and momentum are
 \begin{equation}
P^i=\frac{1}{16 \pi} \int\!\!\!\int{ {{\Lambda}^{i0\alpha m}}_{,m}
n_{\alpha}\,dS}
\label{eqks7}
 \end{equation}
and the physically interesting components
of $J^{ik}$ are
 \begin{equation}
J^{\alpha \beta}=\frac{1}{16 \pi} \int\!\!\!\int{
\left(x^{\alpha}{{\Lambda}^{ \beta0 \sigma m}}_{,m}
-x^{\beta}{{\Lambda}^{ \alpha0 \sigma m}}_{,m}
+\Lambda^{ \alpha 0 \sigma\beta}
\right)n_{\sigma}\,dS}.
\label{eqks8}
 \end{equation}

\section{The Kerr-Newman metric}
\label{sec:metric}

The KN spacetime, characterized by mass, charge, and rotation
parameters, in KS Cartesian coordinates is given by the line
element in Eq.~$(\ref{eqks1})$ with the following choices for $V$
and $l_i$\cite{ref22}:
 \begin{eqnarray}
V &=& \frac{2M\rho^3-
Q^2\rho^2}{2\left(\rho^4+a^2z^2\right)},\label{eq36a}\\
l_i\, dx^i &=& dt+\frac{z}{\rho}\,dz +
\frac{\rho}{\rho^2+a^2} \left(x\,dx +y\,dy\right)
 - \frac{a}{\rho^2+a^2} \left(x\,dy -y\,dx\right),\label{eq36b}
 \end{eqnarray}
where $\rho$ is defined by the positive root of
 \begin{equation}
\frac{x^2+y^2}{\rho^2+ a^2}+ \frac{z^2}{\rho^2}=1.
\label{eq37}
 \end{equation}
$a = 0 $ and $Q = 0$ in
$(\ref{eq36a})$ and $(\ref{eq36b})$ give the RN and Kerr
spacetimes, respectively.
For each value of $\rho$ Eq.~$(\ref{eq37})$ defines a surface
that can be
parameterized by two angular coordinates, $(\theta,\varphi)$, as
follows:
 \begin{eqnarray}
x&=&\sqrt{\rho^2+a^2}\sin\theta\cos\varphi, \nonumber\\
y&=&\sqrt{\rho^2+a^2}\sin\theta\sin\varphi, \nonumber\\
z&=&\rho\cos\theta.
\label{eq38}
 \end{eqnarray}
The components of the outward unit normal vector over the
surface given by $(\ref{eq37})$ (with $\rho$ constant) are
 \begin{eqnarray}
n_x&=&\frac{\rho}{\Upsilon}\sin\theta\cos\varphi, \nonumber\\
n_y&=&\frac{\rho}{\Upsilon}\sin\theta\sin\varphi, \nonumber\\
n_z&=&\frac{\sqrt{\rho^2 + a^2}}{\Upsilon}\cos\theta,
\label{eq39}
 \end{eqnarray}
and the infinitesimal surface element is
 \begin{equation}
dS=\sqrt{\rho^2+a^2}\,\Upsilon\sin\theta\, d\theta\, d\varphi,
\label{eq40}
 \end{equation}
where
 \begin{equation}
\Upsilon=\sqrt{\rho^2+a^2\cos^2\theta}.
\label{eq41}
 \end{equation}

Now, we calculate the energy, momentum, and angular
momentum for the KN metric in KS
cartesian coordinates and use the results of the previous section.
The intermediate mathematical expressions are
very lengthy and therefore we give only the final results, which have
been obtained and checked by means of two different computer algebra
systems. The energy and momentum inside a surface given by
$(\ref{eq37})$ with constant $\rho$ in all the prescriptions of ETLLPW
are
 \begin{eqnarray}
E&=& M-\frac{Q^2}{4 \rho}
\left[
 1 + \frac{\left(a^2+\rho^2\right)}
              { a \rho } \arctan\left(\frac{a}{\rho}\right)
\right], \nonumber\\
P_1&=& P_2= P_3=0.
\label{eq42}
 \end{eqnarray}
The physically important components of the angular momentum in all the
prescriptions of LLPW are
 \begin{eqnarray}
J^{12}&=&a \left\{
 M - \frac{Q^2}{4\rho}
\left[
1 - \frac{\rho^2}{a^2} + \frac{\left(a^2+\rho^2\right)^2}{a^3 \rho}
\arctan\left(\frac{a}{\rho}\right)
\right]
\right\}, \nonumber\\
 J^{23}&=& J^{31}= 0.
\label{eq43}
 \end{eqnarray}

The total energy, momentum, and angular  momentum ($\rho$ approaching
infinity in the above expressions) for the KN metric are $E=M$,
$P_1=P_2=P_3=0$, $J^{12}= M a$ and $J^{23}=J^{31}= 0$. The energy,
momentum, and energy current density components of the ETLLPW
pseudotensors for the KN metric are
 \begin{eqnarray}
\Theta_0{}^0&=& {\cal{T}}_0{}^0=L^{00}=\Sigma^{00}=W^{00}
=\left(\rho^4+2a^2\rho^2-a^2z^2\right) A,\nonumber\\
\Theta_0{}^1&=& - \Theta_1{}^0={\cal{T}}_0{}^1=- {\cal{T}}_1{}^0
 =L^{10}=\Sigma^{10}=W^{10}
=  -2a y \rho^2 A,\nonumber\\
\Theta_0{}^2&=& - \Theta_2{}^0={\cal{T}}_0{}^2=- {\cal{T}}_2{}^0
 =L^{20}=\Sigma^{20}=W^{20}
= 2a x \rho^2 A,\nonumber\\
\Theta_0{}^3&=& \Theta_3{}^0={\cal{T}}_0{}^3= {\cal{T}}_3{}^0
 =L^{30}=\Sigma^{30}=W^{30}=0,
\label{eq44}
 \end{eqnarray}
where
 \begin{equation}
A= \frac{Q^2\rho^4}{8\pi\left(\rho^4+a^2z^2\right)^3}.
\label{eq45}
 \end{equation}
For the Kerr metric $(Q
= 0)$ all the components given in $(\ref{eq44})$ are zero.

\section{The Bonnor-Vaidya metric}
\label{sec:BV}

Two of the present authors (Chamorro and KSV)\cite{ref19} considered
the BV metric in the prescriptions of Einstein and LL and got the same
and reasonable result for the energy distribution. They also got the
same result for the energy current density components.
In the light of
G\"urses and G\"ursey's result\cite{refgg} one sees the reason for the
coincidences.
The BV metric
in KS Cartesian coordinates is given by the line element in
Eq.~$(\ref{eqks1})$ with the following choices for $V$ and $l_i$:
 \begin{eqnarray}
V &=& \frac{M\left(u\right)}{r} -
     \frac{Q^2\left(u\right)}{2r^2},\label{eq48a}\\
l_i\, dx^i &=& dt-dr,\label{eq48b}
 \end{eqnarray}
where $r= \sqrt{x^2+y^2+z^2}$.
The mass and charge parameters, $M(u)$ and
$Q(u)$, depend on the retarded time coordinate $u$ ($u = t - r$). The
pseudotensors of Einstein and LL give the same result\cite{ref19}:
 \begin{equation}
E=M\left(u\right) - \frac{Q^2\left(u\right)}{ 2 r}.
\label{eq50}
 \end{equation}
It is  of interest to note that the Penrose definition also leads to
the same result for the BV metric\cite{ref13}. Next we give the
energy, momentum, and energy current density components for the BV
metric in the prescriptions of ETLLPW.
 \begin{eqnarray}
 \Theta_0{}^0&=& {\cal{T}}_0{}^0=L^{00}=\Sigma^{00}=W^{00}
 =\frac{Q^2}{8 \pi r^4} + r \Delta,\nonumber\\
 \Theta_0{}^1&=& - \Theta_1{}^0={\cal{T}}_0{}^1=- {\cal{T}}_1{}^0
  =L^{10}=\Sigma^{10}=W^{10} =x \Delta,\nonumber\\
 \Theta_0{}^2&=& - \Theta_2{}^0={\cal{T}}_0{}^2=- {\cal{T}}_2{}^0
  =L^{20}=\Sigma^{20}=W^{20}   =y \Delta,\nonumber\\
 \Theta_0{}^3&=& -  \Theta_3{}^0={\cal{T}}_0{}^3=- {\cal{T}}_3{}^0
  =L^{30}=\Sigma^{30}=W^{30}=z \Delta,
 \label{eq51}
 \end{eqnarray}
where
 \begin{equation}
\Delta  =\frac{ Q \dot{Q} - r \dot{M}}
            {4 \pi r^4}.
\label{eq52}
 \end{equation}
The dot over $Q$ and $M$ stands for the derivative with respect to the
retarded time coordinate $u$.

\section{Discussion}
\label{sec:discussion}

The subject of the energy-momentum localization in general relativity
has been debated since the beginning of relativity and it still
continues (for instance, see \cite{ref23}). Bondi\cite{ref23} argued
that a non-localizable form of energy is inadmissible in relativity
and so its location can in principle be found. Following the Einstein
pseudotensor, a large number of coordinate-dependent as well as
coordinate independent definitions of energy, momentum, and angular
momentum in general relativity have been given in the literature.
There is no adequate coordinate-independent prescription for
energy-momentum localization in general relativity.
Bergqvist\cite{ref10} investigated seven different definitions of
energy and reported that no two definitions give the same result for
the RN and Kerr spacetimes. The well known quasi-local definition for
energy, momentum, and angular momentum given by Penrose, which gave
reasonable result for several spacetimes, has not succeeded to handle
the Kerr metric\cite{ref13}.

In the present paper we have obtained the energy and angular momentum
for the KN metric for arbitrary values of the mass, charge, and
rotation parameters. The pseudotensors of ETLLPW give the same and
reasonable energy distribution. Again the symmetric pseudotensors of
LLPW give the same and reasonable angular momentum distribution for
this metric. They also give the same energy and energy current density
components for the KN metric. For the KN black hole the energy is
distributed by its interior as well as exterior whereas for the Kerr
black hole the energy is confined to its interior. This proves a
previous conjecture of KSV\cite{ref16} and is compatible with
Cooperstock's conjecture\cite{refCo}. It is clear from
$(\ref{eq42})$ and $(\ref{eq43})$ that the energy distribution is
independent of the sign on the charge as well as rotation parameters
whereas the  direction of the angular momentum  depends on the sign of
the rotation parameter and is independent of the sign on the charge
parameter. This is obviously a convincing result. The total energy and
angular momentum ($\rho$ approaching infinity in $(\ref{eq42})$ and
$(\ref{eq43})$) are $M$ and $M a$, respectively. For the RN  metric
($a = 0$), one gets
 \begin{equation}
E=M-\frac{Q^2}{2 r}.
\label{eq46}
 \end{equation}
The definitions of Penrose as well as that of Hayward give the same
result for the RN metric \cite{ref12,ref24}. Also, for the BV metric
the pseudotensors of ETLLPW give the same result (see (\ref{eq50})) as
given by the Penrose definition\cite{ref13}.

Summarizing, the energy-momentum localization has been a longstanding
``recalcitrant problem'' in general relativity. Despite many
painstaking efforts no adequate coordinate-independent definition is
known. We have shown that several pseudotensors give the same and
reasonable result for the KN as well as the BV spacetimes when
calculations are carried out in KS Cartesian coordinates. Different
pseudotensors giving the same results for local quantities (in KS
Cartesian coordinates) does not seem to be accidental. It could be of
interest to investigate this problem further.

\acknowledgments

One of us (KSV) thanks the Basque Government
and the Tata Institute of Fundamental
Research for financial support and
J.\ M.\ Aguirregabiria, A.\ Chamorro, I.\ Egusquiza, and M.\ Rivas for
hospitality. This work was supported in part by the University of the
Basque Country under contract UPV/EHU 172.310-EA034/94. We thank
F.\ I.\ Cooperstock for pointing out that a previous version of this
work could be placed in perspective by properly taking into account
the work by G\"urses and G\"ursey\cite{refgg}.
We also thank R.\ Jackiw, J.\ Katz, A.\ Komar, and K.\ P.\ Tod
for helpful correspondence.

\end{document}